\documentclass[runningheads]{llncs}

\usepackage{graphicx}

\input epsf

\begin{document}
\pagestyle{headings}
\mainmatter

\title{Automatic adaptor synthesis for protocol transformation}
\titlerunning{Automatic adaptor synthesis}

\author{Marco Autili  \and Paola Inverardi \and Massimo Tivoli}
\authorrunning{Marco Autili et al.}

\institute{University of L'Aquila\\
Dip. Informatica\\
fax: +390862433057\\
via Vetoio 1, 67100 L'Aquila\\
\email{\{marco.autili, inverard, tivoli\}@di.univaq.it}}

\maketitle

\begin{abstract}
Adaptation of software components is an important issue in
Component Based Software Engineering (CBSE). Building a system
from reusable or \emph{Commercial-Off-The-Shelf} (COTS) components
introduces a set of issues, mainly related to compatibility and
communication aspects. Components may have incompatible
interaction behavior. Moreover it might be necessary to enhance
the current communication protocol to introduce more sophisticated
interactions among components. We address these problems enhancing
our architectural approach which allows for detection and recovery
of integration mismatches by synthesizing a suitable coordinator.
Starting from the specification of the system to be assembled and
from the specification of the needed protocol enhancements, our
framework automatically derives, in a compositional way, the glue
code for the set of components. The synthesized glue code avoids
interaction mismatches and provides a protocol-enhanced version of
the composed system.
\end{abstract}

\section{Introduction} \label{introduction}

Adaptation of software components is an important issue in
Component Based Software Engineering (CBSE). Nowadays, a growing
number of systems are built as composition of reusable or
\emph{Commercial-Off-The-Shelf} (COTS) components. Building a
system from reusable or from COTS components introduces a set of
problems, mainly related to communication and compatibility
aspects~\cite{Szy98}. In fact, components may have incompatible
interaction behavior~\cite{Gao95}. Moreover, it might be necessary
to enhance the current communication protocol to introduce more
sophisticated interactions among components. These enhancements
(i.e. protocol transformations) might be needed to achieve
dependability, to add extra-functionality and/or to properly deal
with system's architecture updates (i.e. component aggregating,
inserting, replacing and removing).

By referring to~\cite{wrappers_formalization}, many ad-hoc
solutions can be applied to enhance dependability of a system.
Each solution is formalized as a process algebra specification of
communication protocol enhancements. While this approach provides
a formal specification for a useful set of protocol enhancements,
it lacks in automatic support in applying the specified
enhancements.

In this paper, by referring to~\cite{wrappers_formalization}, we
exploit and improve an existent
approach~\cite{bertinoro,tosem,cbse7} for automatic synthesis of
failure-free coordinators for COTS component-based systems. The
existent approach automatically synthesizes a coordinator to
mediate the interaction among components by avoiding incompatible
interaction behavior. This coordinator represents a starting glue
code. In this paper we propose an extension that makes the
coordinator synthesis approach also able to automatically
transform the coordinator's protocol by enhancing the starting
glue code. These enhancements might be performed to achieve
dependability or to implement more complex interactions or to deal
with system's architectural updates. This, in turn, allows us to
introduce extra-functionality such as fault-tolerance,
fault-avoidance, security, load balancing, monitoring and error
handling into composed system. A modified version of the approach
described in this paper has been applied to a more specific
application domain than the general domain we consider. This is
the domain of reliability enhancement in component-based
system~\cite{tivgardaland}. Starting from the specification of the
system to be assembled, and from the specification of the protocol
enhancements, our framework automatically derives, in a
compositional way, the glue code for the set of components. The
synthesized glue code avoids interaction mismatches and provides a
protocol-enhanced version of the composed system. By assuming i) a
behavioral description of the components and of the coordinator
forming the system to be assembled and ii) a specification of the
protocol enhancements needed on the coordinator, we automatically
derive a set of new coordinators and extra-components to be
assembled with the old coordinator in order to implement the
specified enhanced protocol. Each extra-component is synthesized
as a wrapper. A wrapper component intercepts the interactions
corresponding to the old coordinator's protocol in order to apply
the specified enhancements. It is worthwhile noticing that we
might use existent third-party components as wrappers. In this
case we do not need to synthesize them. The new coordinators are
needed to assemble the wrappers with the old coordinator and the
rest of the components forming the composed system.

Our approach is compositional in the automatic synthesis of the
enhanced glue code. That is each enhancement is performed as a
modular protocol transformation. This allows us to perform a
protocol transformation as composition of other protocol
transformations by impacting on the reusability of the synthesized
glue code. For example if we add a new component or we
replace/remove an existent component, we enhance the existent glue
code by completely reusing it and by synthesizing a new modular
glue code to be assembled with the existent one. The
compose-ability, in turn, improves the space complexity of the old
coordinator synthesis approach~\cite{bertinoro,tosem,cbse7}.

The paper is organized as follows. Section~\ref{background}
introduces background notions helpful to understand our approach.
Section~\ref{methoddescription} illustrates the technique
concerning with the enhanced coordinator synthesis.
Section~\ref{conclusion} discusses future work and concludes.

\section{Background} \label{background}

In this section, we provide the background needed to understand
the approach illustrated in Section~\ref{methoddescription}.

\subsubsection{\textbf{The reference architectural style:}}
\label{CBA_architecture}

\begin{figure}[h]
\centerline{\epsfxsize=3.5cm \epsfysize=1.5cm
\epsfbox{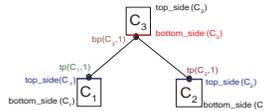}} \caption{An architecture
sample} \label{arch_sample}
\end{figure}

The architectural style we refer to, consists of a components and
connectors style. Components define a notion of top and bottom
side. The top (bottom) side of a component is a set of top
(bottom) ports. Connectors are synchronous communication channel
which define a top and a bottom role. A top (bottom) port of a
component may be connected to the bottom (top) role of a single
connector. Components communicate synchronously by passing two
types of messages: notifications and requests. A notification is
sent downward, while a request is sent upward. A top-domain of a
component is the set of requests sent upward and of received
notifications. Instead a bottom-domain is the set of received
requests and of notifications sent downward. This style is a
generic layered style. Since it is always possible to decompose a
$n$-layered system in $n$ single-layered sub-systems, in the
following of this paper we will only deal with single layered
systems. Refer to~\cite{bertinoro} for a description of the above
cited decomposition. In Figure~\ref{arch_sample}, we show a sample
of a software architecture built by using our reference style.

$bp(C_j,k)$ is the bottom port $k$ of component $C_j$. $tp(C_j,h)$
is the top port $h$ of component $C_j$.

\subsubsection{\textbf{Configuration formalization:}}
\label{configuration_formalization}

\begin{figure}[h]
\centerline{\epsfxsize=3.5cm \epsfysize=2.5cm
\epsfbox{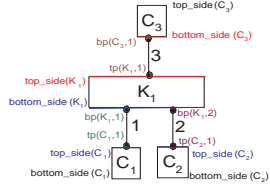}} \caption{A coordinator based
architecture sample} \label{cba_arch_sample}
\end{figure}

We consider a particular configuration of the composed system
which is called \emph{Coordinator Based Architecture} (CBA). It is
defined as \emph{a set of components directly connected, through
connectors, in a synchronous way to one or more coordinators}. It
is worthwhile noticing that we use the style described in
Section~\ref{CBA_architecture} to build a CBA. That is, a
coordinator in a CBA is implemented as a component which is
responsible only for the routing of messages among the others
components. Moreover, the coordinator exhibits a strictly
sequential input-output behavior\footnote{Each input action is
strictly followed by the corresponding output action.}. All
components (i.e. coordinators too) and system behaviors are
specified and modelled as Labelled Transition Systems (LTSs). This
is a reasonable assumption because from a standard incomplete
behavioral specification (like \emph{Message Sequence Charts}
specification) of the coordinator-free composed system we can
automatically derive these LTSs by applying the old coordinator
synthesis approach. Refer to~\cite{bertinoro,tosem,cbse7}, for
further details. In Figure~\ref{cba_arch_sample}, we show a sample
of a coordinator based software architecture built by using our
reference style.

$K_1$ is the coordinator component.

\section{Method description} \label{methoddescription}

In this section, for the sake of brevity, we informally describe
our method. Refer to~\cite{autili:tesi} for a formal description
of the whole approach.

The problem we want to face can be informally phrased as follows:
\emph{given a CBA system $S$ for a set of black-box components
interacting through a coordinator $K$ and a set of coordinator
protocol enhancements $E$ automatically derive the corresponding
enhanced CBA system $S^{\prime}$}.

\begin{figure}[h]
\centerline{\epsfxsize=11cm \epsfysize=5.5cm
\epsfbox{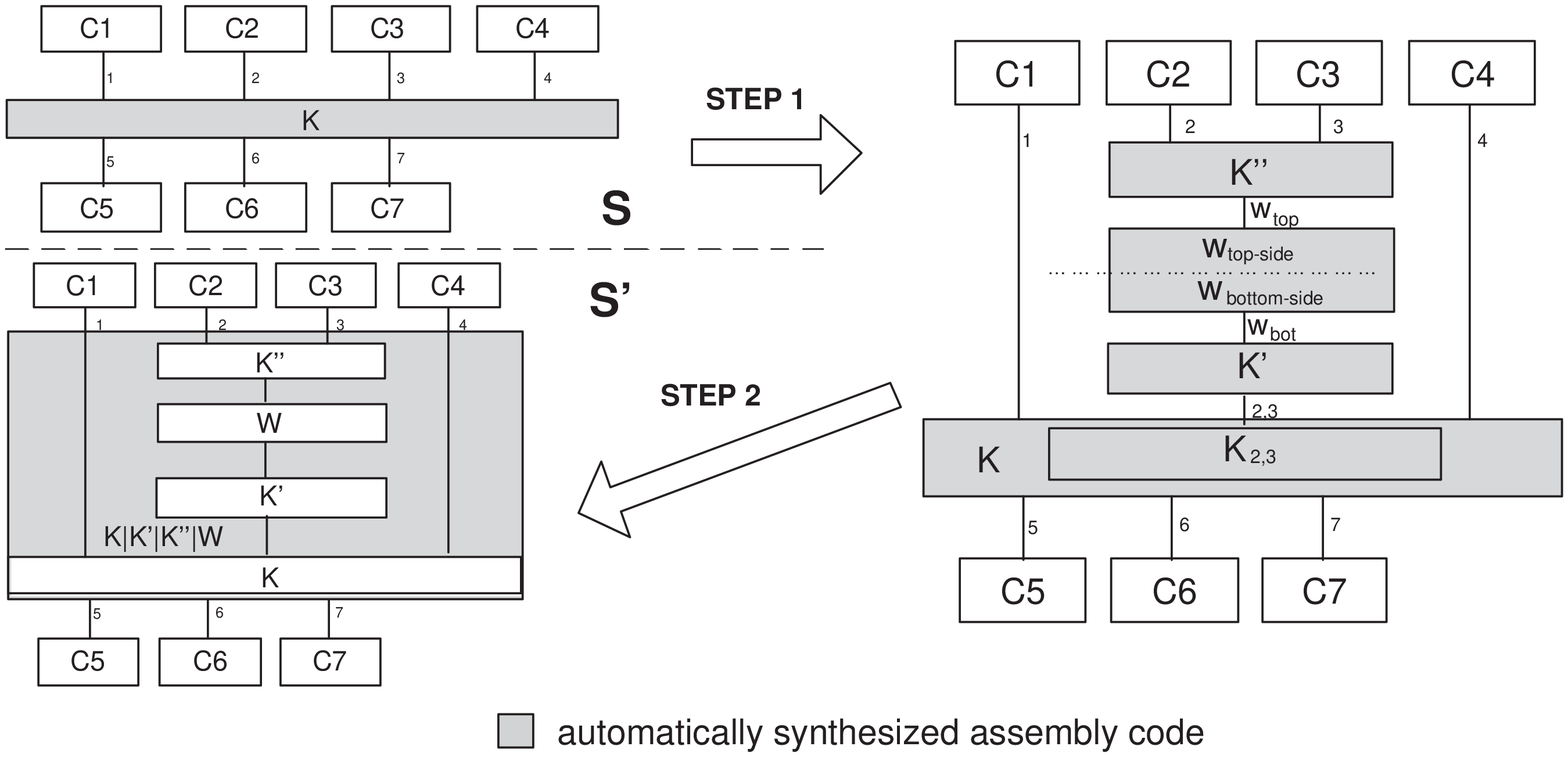}} \caption{2 step method}
\label{fig:figura1}
\end{figure}

We are assuming that a specification of the CBA system to be
assembled is provided in terms of a description of components and
coordinator behavior as LTSs. Moreover we assume that a
specification of the coordinator protocol enhancements to be
applied exists. This specification is given in form of \emph{basic
Message Sequence Charts} (bMSCs) and \emph{High level MSCs}
(HMSCs) specifications~\cite{itu:msc}. In the following, we
discuss our method proceeding in two steps as illustrated in
Figure~\ref{fig:figura1}.

In the first step, by starting from the specification of the CBA
system, we apply the specified coordinator protocol enhancements
to derive the enhanced version of CBA. We recall that we apply
coordinator protocol enhancements by inserting a wrapper component
between the coordinator in the CBA (i.e. $K$ of
Figure~\ref{fig:figura1}) and the portion of composed system
concerned with the coordinator protocol enhancements (i.e. the set
of $C2$ and $C3$ components of Figure~\ref{fig:figura1}). It is
worthwhile noticing that we do not need to consider the entire
coordinator model but we just consider the
\emph{"sub-coordinator"} which represents the portion of $K$ that
communicates with the components $C2$ and $C3$ (i.e. the
\emph{"sub-box"} $K_{2,3}$ of Figure~\ref{fig:figura1}). The
wrapper component intercepts the messages exchanged between
$K_{2,3}$, $C2$ and $C3$ and applies the specified enhancements on
the interactions performed on the communication channels 2 and 3
(i.e. connectors 2 and 3 of Figure~\ref{fig:figura1}). We first
decouple $K$, $C2$ and $C3$ to ensure that they are no longer
directly synchronized. This is done by decoupling $K_{2,3}$, $C2$
and $C3$. Then we automatically derive a behavioral model of the
wrapper component (i.e. a LTS) from the bMSCs and HMSCs
specification of the coordinator protocol enhancements. We do this
by exploiting our implementation of the technique described
in~\cite{implied_scenarios} and also used in the old coordinator
synthesis approach~\cite{bertinoro,tosem,cbse7,invtiv:monterey}.
Finally, the wrapper is interposed between the \emph{top-side} of
$K_{2,3}$ and the \emph{bottom-side} of $C_2$ and $C_3$ by
automatically synthesizing two new coordinators $K^\prime$ and
$K^{\prime\prime}$. We do this by exploiting the coordinator
synthesis approach formalized and developed in~\cite{autili:tesi}.

In the second step, we derive the implementation of the
synthesized glue code formed by composing the wrapper component,
with the old and new synthesized coordinators (i.e. $W$, $K$,
$K^{\prime\prime}$ and $K^{\prime}$ of Figure~\ref{fig:figura1}
respectively). This code represents the coordinator in the
enhanced version $S^{\prime}$ of the CBA system $S$ (i.e. the
coordinator $(K \mid K^{\prime} \mid K^{\prime\prime} \mid W)$ in
system $S^{\prime}$ of Figure~\ref{fig:figura1}). By referring to
Figure~\ref{fig:figura1}, the enhanced coordinator $(K \mid
K^{\prime} \mid K^{\prime\prime} \mid W)$ in $S^{\prime}$ may be
treated in the same way of the old coordinator $K$ in $S$ with
respect to the application of the new coordinator protocol
enhancements. This allows us to achieve compose-ability of
different coordinator protocol enhancements (i.e. different
protocol's transformations). In other words, our approach is
compositional in the automatic synthesis of the enhanced glue
code.

\subsection{Application example}

\begin{figure}[h]
    \centerline{\epsfxsize=8.0cm \epsfysize=3.0cm
    \epsfbox{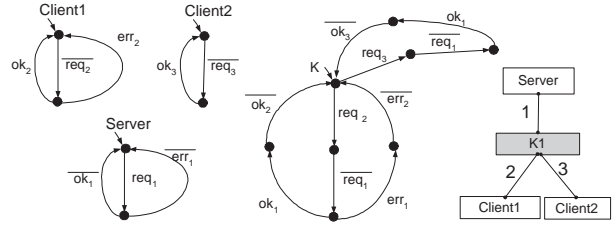}} \caption{A Client-Sever CBA system and its specification}
    \label{fig:K}
    \end{figure}

\begin{figure}[h]
    \centerline{\epsfxsize=9.0cm \epsfysize=4.5cm
    \epsfbox{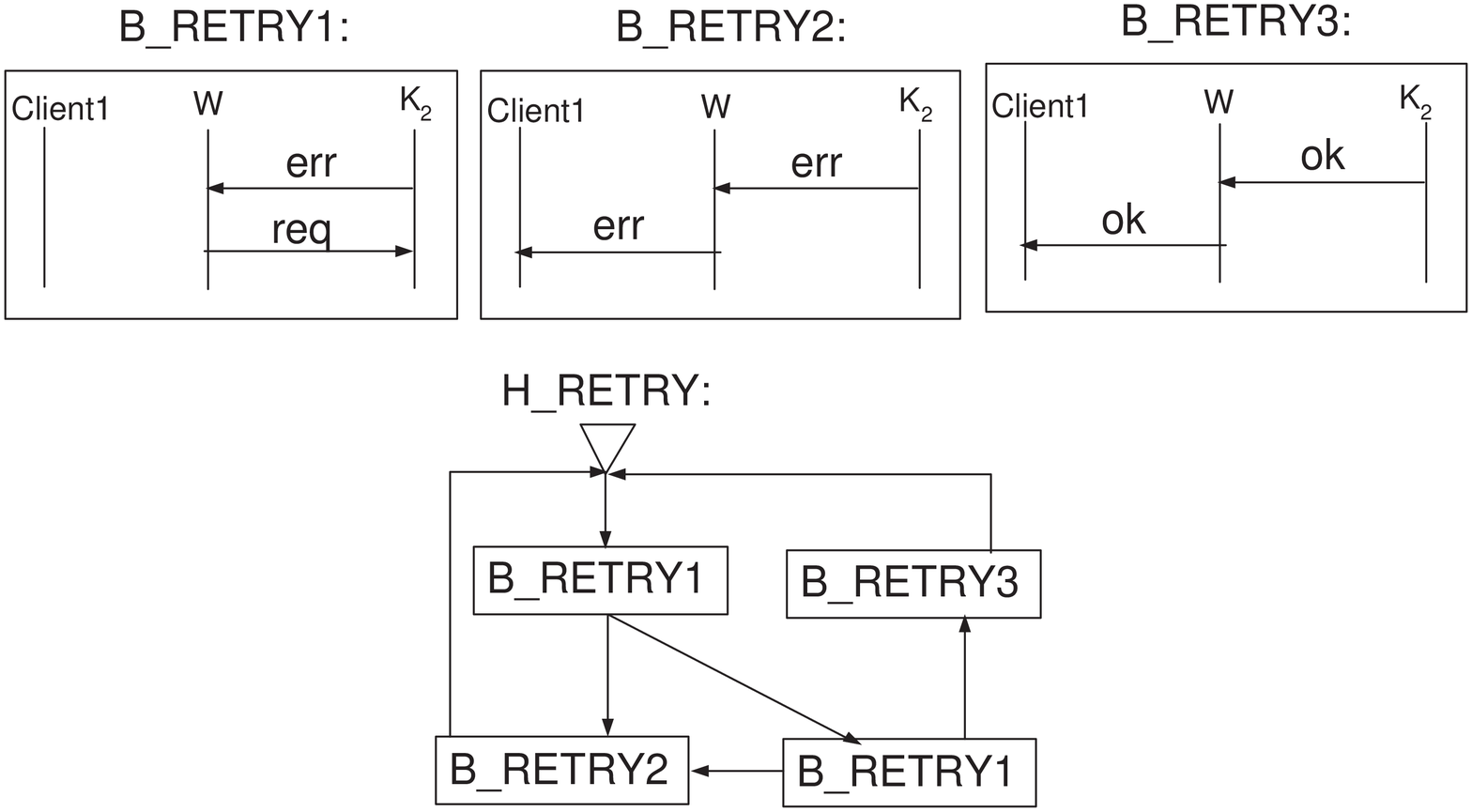}} \caption{The $bMSCs$ and the $hMSC$ of the coordinator protocol
    enhancements: \emph{retry} policy} \label{fig:retry}
    \end{figure}

    In this section, by means of an explanatory example, we show the processed steps of our method.
    In Figure ~\ref{fig:K}, we consider a CBA system and its specification.
    Moreover in Figure ~\ref{fig:retry}, we consider the specification of the coordinator protocol enhancements.

    $Client1$ performs a request and waits for a response: erroneous or successful\footnote{The transitions labelled with
    $\alpha_i$ denote input actions while the transitions labelled with $ \bar{\alpha_i} $ denote output
    actions on the communication channel \emph{i}(i.e. the connector \emph{i}).}.
    $Server$ may answer with an error message
    (the error could be either due to an upper-bound on the number of request $Server$ can accept
    simultaneously or due to a general transient-fault). Now, let $Client1$ be
    an interactive client and once an error message occurs, it shows a
    dialog window displaying information about the error. The user might not appreciate this error message
    and he might lose the degree of trust in the system.
    By recalling that the dependability of a system reflects the user's
    degree of trust in the system, this example shows a commonly practiced
    dependability-enhancing technique. The wrapper attempts to hide
    the error to the user by re-sending the request a finite
    number of times. This is the \emph{retry} policy specified in
    Figure ~\ref{fig:retry}. The wrapper W re-sends at most two
    times.

    \begin{figure}[h]
    \centerline{\epsfxsize=2.5cm \epsfysize=2.0cm
    \epsfbox{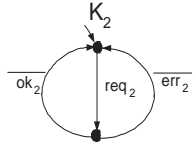}} \caption{\emph{Bottom domain Actual Behavior
    Graph, $KBAC_{2}$, of the $K1$ coordinator}} \label{fig:KBAC}
    \end{figure}

%
%

%

\begin{figure}[h]
    \centerline{\epsfxsize=11cm \epsfysize=5.5cm
    \epsfbox{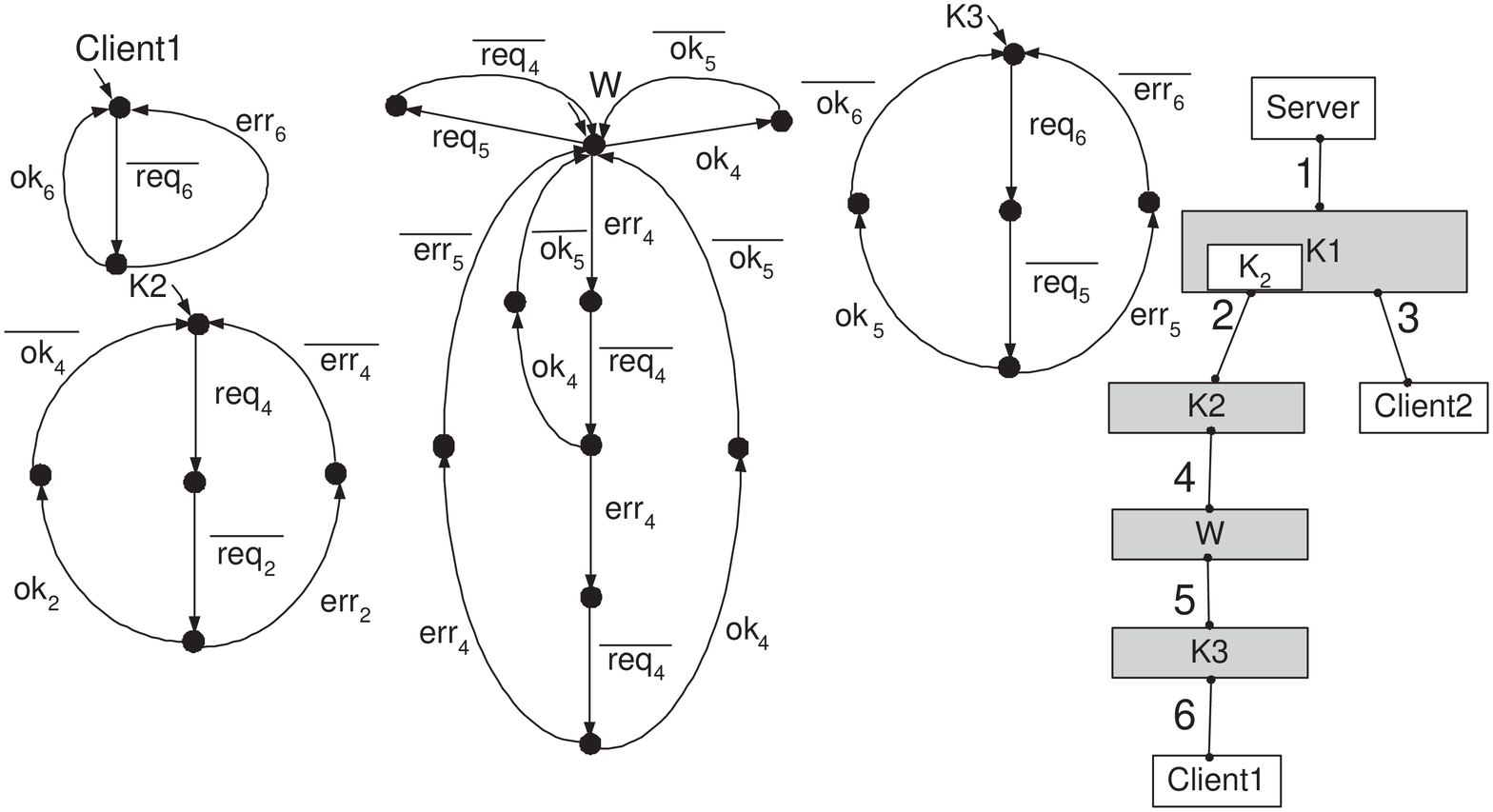}} \caption{Enhanced version of the \emph{Client-Server} system.} \label{fig:K1}
    \end{figure}

    $K_2$ is the ``\emph{sub-coordinator}'' which
    represents the portion of K1 communicating with $Client1$.
    Its ``\emph{real}'' behavior is described by the so called \emph{Bottom domain Actual Behavior Graph
    of the $K1$ coordinator, $KBAC_{2}$, restricted to the Client1 component} (the LTS in Figure
    ~\ref{fig:KBAC}). This graph is obtained by filtering the LTS specification
    of K1. This is done by using the algorithm formalized in
    ~\cite{autili:tesi}. Then we automatically derive a behavioral model of
 $K_2$ (i.e. a LTS) from the bMSCs and HMSCs specification of
the \emph{retry} police. If this model differs from the
``\emph{real}'' behavior then the enhancement cannot be performed
because its specification does not ``\emph{reflect}'' the behavior
of K1; otherwise, we first automatically derive the LTS
specification of the wrapper W and then we insert it between K
(i.e. $K_2$) and the top-side of $Client1$. This is done by
automatically synthesizing two new coordinators: K2 and K3 (Figure
~\ref{fig:K1}). The wrapper inserting procedure is formalized in
~\cite{autili:tesi}.

\section{Conclusion and future work} \label{conclusion}

In this paper we propose and briefly describe an extension of the
approach presented in~\cite{bertinoro,tosem,cbse7}. The extension
is performed to make the old approach able to achieve
compose-ability and to deal with both problems in the area of
dependability enhancement and problems raised by system's
architecture updates. We have formalized the whole approach. Refer
to~\cite{autili:tesi} for details about the formalization of the
approach.

The key results are: i) the extended approach is compositional in
the automatic synthesis of the enhanced coordinator, ii) by
achieving compose-ability, we improve the space complexity of the
old coordinator synthesis approach~\cite{bertinoro,tosem,cbse7}
and iii) the enhanced coordinator is adequate with respect to
\emph{data translation, components inserting, removing and
replacing, monitoring, error handling, security, dependability
enhancement}.

The automation and applicability of the old coordinator synthesis
approach~\cite{bertinoro,tosem} is supported by our tool called
\emph{"SYNTHESIS"}~\cite{cbse7}. As future work, we plan to: i)
extend the current implementation of the \emph{"SYNTHESIS"} tool
to support the automation of the extended coordinator synthesis
approach~\cite{autili:tesi} presented in this paper; ii) think
about a more user-friendly and real-scale context specification of
the coordinator protocol enhancements (e.g. UML2 Interaction
Overview Diagrams and Sequence Diagrams); iii) validate the
applicability of the whole approach to real-scale examples.

\bibliographystyle{abbrv}
\bibliography{wcat04}

\end{document}